\documentclass[12pt]{article}
\usepackage[utf8]{inputenc}
\usepackage{multirow}
\author{Author Name}
\DeclareUnicodeCharacter{03A6}
{\ensuremath{\Phi}}

\usepackage{amsmath}

\usepackage{amssymb}
\usepackage{tabularx}
\usepackage{newunicodechar}
\newunicodechar{ }{~}

\usepackage{doi}

\usepackage{longtable}

\pdfoutput=1
\usepackage{graphicx}

\usepackage{float} 

\usepackage{cite}
\usepackage{tabularx}

\usepackage{microtype} 

\usepackage[utf8]{inputenc}
\usepackage{longtable}
\usepackage{array}
\usepackage{ragged2e}
\usepackage[utf8]{inputenc}
\usepackage{enumitem}
\usepackage{xurl}

\usepackage{float}

\title{The Case for a Horizontal Federated AI Operating System for Telcos}
\author{Sebastian Barros}
\date{June 8th , 2025}  

\begin{document}
\renewcommand{\arraystretch}{1.2}

\maketitle

\begin{abstract}
As artificial intelligence capabilities rapidly advance, Telco operators face a growing need to unify fragmented AI efforts across customer experience, network operations, and service orchestration. This paper proposes the design and deployment of a horizontal federated AI operating system tailored for the telecommunications domain. Unlike vertical vendor-driven platforms, this system acts as a common execution and coordination layer, enabling Telcos to deploy AI agents at scale while preserving data locality, regulatory compliance, and architectural heterogeneity.

We argue that such an operating system must expose tightly scoped abstractions for telemetry ingestion, agent execution, and model lifecycle management. It should support federated training across sovereign operators, offer integration hooks into existing OSS and BSS systems, and comply with TM Forum and O-RAN standards. Importantly, the platform must be governed through a neutral foundation model to ensure portability, compatibility, and multi-vendor extensibility.

This architecture offers a path to break current silos, unlock ecosystem-level intelligence, and provide a foundation for agent-based automation across the Telco stack. The case for this horizontal layer is not only technical but structural, redefining how intelligence is deployed and composed in a distributed network environment.
\end{abstract}

\section{Introduction}

The telecommunications industry is undergoing a transformation driven by the convergence of artificial intelligence, software-defined networking, and cloud-native architecture. As networks evolve into data-intensive, programmable systems, Telco operators are increasingly exploring AI to optimize performance, predict anomalies, and enhance customer experience. However, most AI initiatives within Telcos remain fragmented, domain-specific, and reliant on proprietary vendor platforms.

This vertical fragmentation prevents AI models from achieving cross-domain interoperability or composability. Models developed for network optimization are often disconnected from customer service insights or operational workflows. Unlike hyperscalers, which operate unified platforms for deploying AI at scale, Telcos lack a shared execution layer that enables coordination across distributed systems. Each new use case requires point-to-point integration and repeated infrastructure provisioning, creating unnecessary duplication and complexity.

To address this gap, we propose a federated AI operating system for Telcos: a horizontal software substrate that exposes standardized abstractions for model execution, telemetry ingestion, agent lifecycle management, and integration with existing OSS and BSS systems. This operating system is designed to enable a new class of interoperable AI agents capable of operating across diverse Telco environments while preserving data sovereignty and regulatory compliance.

The system supports federated orchestration and training, allowing operators to collaborate without centralizing sensitive datasets. It also provides compatibility layers with existing standards from TM Forum, O-RAN, and GSMA initiatives, ensuring that new AI workloads can coexist with current operational infrastructure. Governance, certification, and open interfaces are essential to prevent fragmentation and support ecosystem-wide collaboration.

This paper outlines the architecture, technical design, federation mechanisms, agent models, and governance strategy required to realize this vision. It presents a roadmap for creating a neutral, AI-native layer that abstracts the Telco stack and enables scalable, secure, and extensible intelligence across the global communications infrastructure.

\section{The Structural Problem in Telco AI}

\subsection{AI in Telecommunications is Fragmented by Design}

Despite widespread interest in artificial intelligence in telecommunications, real-world deployments remain siloed, confined to specific domains like RAN fault detection, BSS churn modeling, or transport anomaly identification. GSMA Intelligence reports that while 65 percent of operators are piloting AI, few have integrated it across multiple operational domains, reflecting a fragmented deployment landscape \cite{gsma_intel_2024}.

The underlying cause is the legacy structure of telecom systems: vertically integrated stacks for network management, billing, customer care, and more, each managed by different vendors. This architecture contrasts with cloud and enterprise IT, which use shared APIs, microservices, and open‑source frameworks. The consequence is that AI models built for one vertical are incompatible and non-portable across others, severely limiting scale and reuse.

\subsection{Vendor Lock-In and Proprietary Model Silos}

Most AI capabilities in Telcos today are supplied through vendor-specific tooling, embedded in OSS/BSS products, cloud platforms, or network hardware. These systems expose closed APIs and retain intelligence within proprietary pipelines. For example, SoftBank’s Large Telecom Model (LTM), while impressive in network operations, remains proprietary and tied to SoftBank’s systems, reducing its applicability across operators \cite{softbank_ltm_2025}.

At the same time, hyperscale AI platforms delivered via the cloud often treat Telcos merely as data sources. Without a neutral, interoperable intelligence layer, Telcos risk becoming commoditized execution platforms rather than creators of domain-specific AI value. This amplifies existing strategic risks around vendor dependency and inhibits cross-domain innovation.

\subsection{Data Sovereignty and Governance Constraints}

Telecom operators manage highly sensitive data, including subscriber identities, call detail records, geolocation logs, and network telemetry. These datasets are regulated by strict national and regional laws, such as GDPR in Europe or data localization requirements in India, Brazil, and the Middle East. As a result, most operators cannot freely aggregate or centralize this data for model training, especially in cross-border scenarios.

This has created a core tension between the need for large-scale learning and the imperative to preserve sovereignty. Recent initiatives like the GSMA Open-Telco LLM Benchmarks explicitly acknowledge this challenge, proposing federated learning as a privacy-preserving architecture that allows Telcos to benchmark and contribute to shared models without moving raw data off-premises \cite{gsma_open_llm_2025}.

In federated learning, each operator trains a model locally and shares only model updates or parameter gradients with a central coordinator. This structure is essential for any AI operating system intended to scale globally while respecting data governance laws.

\subsection{Lack of AI Abstractions and Interfaces}

While telecom standards bodies have made progress in modularization, such as with the O-RAN Alliance’s RIC (RAN Intelligent Controller) and 3GPP’s NWDAF (Network Data Analytics Function), these efforts are still domain-specific. They focus narrowly on the radio or core and do not provide cross-domain intelligence orchestration, nor do they enable composability across service layers, OSS/BSS platforms, and customer-facing applications.

Academic work has explored how to bridge this gap using federated AI components within O-RAN, enabling distributed ML services with privacy and elasticity guarantees \cite{oran_federated_synergies_2023}. More recent research has shown the feasibility of deploying xApps that perform federated learning in O-RAN environments, but these architectures remain in early experimental phases \cite{fedora_oran_2025}.

In contrast, industries such as cloud computing and finance have embraced open AI orchestration layers (e.g., MLflow, TFX, Kubernetes-native AI operators) to streamline deployment and reuse. The telecom industry lacks a comparable horizontal AI abstraction layer that integrates AI services at scale across heterogeneous infrastructure.

\subsection{Structural Summary: Why Telco AI Cannot Scale Today}

The barriers discussed above, data fragmentation, vendor lock-in, regulatory constraint, and the absence of standardized interfaces—are not just implementation challenges. They are architectural voids.

\begin{enumerate}
    \item \textbf{Data is fragmented} across vertical domains and regulatory zones.
    \item \textbf{AI models are proprietary} and locked inside vendor ecosystems.
    \item \textbf{Privacy and sovereignty} restrict cross-border or cloud-native learning.
    \item \textbf{There is no open AI abstraction layer} to enable cross-domain composition and reuse.
\end{enumerate}

These limitations prevent the emergence of scalable, interoperable, and composable AI systems in telecom. Overcoming them requires not another model, but an infrastructure layer: a federated, multi-modal AI operating system horizontally integrated across the Telco stack. We describe this in detail in the next section.

\section{Architectural Vision for a Horizontal Federated AI Operating System}

\subsection{Design Principles}

The proposed AI operating system is founded on a set of design principles intended to overcome the structural limitations outlined in Section 2. First, the system must operate horizontally across telecom domains, encompassing OSS/BSS, transport, RAN, and customer service, rather than being confined to siloed applications. This horizontal approach is critical for enabling AI to function as a shared intelligence substrate rather than as isolated model deployments.

Second, the architecture is federated by design. Rather than aggregating raw data in a centralized cloud, the system supports on-premise model training and shares only encrypted parameter updates or distilled outputs. This ensures regulatory compliance and respects national sovereignty requirements, a growing concern in cross-border telecom AI deployments. Third, the system must support multi-modal learning, integrating text-based models (such as large language models), graph-based models for network topology and fault reasoning, and time-series forecasters for KPIs and predictive maintenance.

In addition, the system promotes composable AI agents. Instead of building monolithic models, developers can construct modular agents, such as a NOC assistant or SLA enforcement engine, that invoke and coordinate multiple underlying models. Finally, the system must support edge-native deployment, enabling inference and training at MEC sites or network edge locations to meet real-time constraints. These principles align with emerging research in federated learning over telecom infrastructure, such as elastic orchestration frameworks for O-RAN-based distributed AI systems \cite{oran_federated_synergies_2023}.

\subsection{Layered Architecture Overview}

The proposed architecture is structured into five primary layers that abstract complexity and promote interoperability. At the base is the infrastructure layer, which includes the radio access network, transport, core, and legacy OSS/BSS systems. Above this lies the data abstraction layer, responsible for ingesting and normalizing diverse datasets, including telemetry, alarms, ticket logs, CRM interactions, and network topology, into a unified schema accessible by AI services.

The model orchestration layer hosts multi-modal learning components. This includes telecom-specific large language models trained on configuration manuals, ticket histories, and operational text; graph neural networks that embed topology information for root cause analysis; and time-series models that forecast traffic or detect KPI anomalies. These models are containerized and callable via standardized APIs.

Above this sits the federated coordination layer, which enables privacy-preserving learning across multiple operators. Local models are trained on each Telco’s infrastructure, and only secure updates are aggregated to refine shared models. This layer may be coordinated by a neutral body or consortium and can include mechanisms for policy enforcement and differential privacy. At the top, the AI abstraction layer exposes SDKs and agent-building interfaces that allow developers to compose telecom-native applications, such as SLA copilots, anomaly responders, or self-healing loop controllers.

\subsection{Federated Intelligence Fabric}

A defining feature of the architecture is its embedded support for federated learning. Rather than depending on centralized data lakes or cloud training, each Telco can retain data on-premise while contributing to a shared model ecosystem. This is particularly relevant in regulated markets, where subscriber data cannot leave national boundaries. Inspired by elastic federated learning frameworks in O-RAN, the system dynamically allocates bandwidth, schedules updates, and adjusts node participation based on network conditions \cite{oran_federated_synergies_2023}.

Recent empirical studies validate this approach. For example, FedORA demonstrated a federated learning resource allocator deployed as a xApp within an O-RAN architecture, showing that it is possible to train and coordinate distributed AI agents within production-grade telecom environments \cite{fedora_oran_2025}. These experiments show that federated learning can be made communication-efficient and real-time responsive, two properties essential for Telco-grade operations.

Together, these layers and design patterns establish a practical and scalable path toward realizing a Telco-native AI operating system, one that is modular, interoperable, privacy-preserving, and vendor-neutral by default.

\subsection{Multi-Modal Intelligence Stack}

A robust AI operating system for telecom must support multiple modalities of intelligence to effectively reason over the heterogeneous data landscape of modern networks. Traditional AI systems in telecom have focused on narrow forecasting or anomaly detection models trained on structured time-series data. However, telecom operations increasingly rely on insights extracted from diverse sources: network topology graphs, customer support transcripts, maintenance logs, alarms, and unstructured policy documentation.

To address this, the AI OS must provide native support for large language models (LLMs), graph neural networks (GNNs), and time-series predictors, orchestrated through a common interface. LLMs fine-tuned on telecom-specific corpora (e.g., ticket histories, network configuration manuals, SLA contracts) can automate NOC workflows, assist with troubleshooting, and contextualize alerts. Graph-based models are particularly well-suited for analyzing complex network topologies and identifying root causes of faults that propagate across nodes or layers, as demonstrated in recent works like TeleNet and GAT-NEQ \cite{telenet_gnn_2024}. Time-series forecasting remains central to predictive maintenance, load balancing, and energy efficiency applications, where temporal dependencies are key.

By converging these capabilities, the system forms a multi-modal AI stack that can be queried by agents, allowing cross-modal reasoning, such as using a time-series anomaly to trigger an LLM-generated summary and a graph-based root cause hypothesis. The orchestration layer must support model versioning, A/B testing, and real-time feedback loops to continuously improve performance across modalities.

\subsection{Edge Deployment and Real-Time Constraints}

One of the most critical architectural considerations in telecom AI systems is the need for real-time inference and decision-making at the edge. Centralized cloud-based inference introduces latency and increases the risk of SLA violations in time-sensitive applications such as dynamic RAN optimization, self-healing loops, or fraud detection in transactional systems.

To meet these constraints, the AI OS must support edge-native deployment through containerized inference runtimes hosted at MEC nodes, regional data centers, or even on-site RIC components. Model compression and distillation techniques can be used to reduce model size while preserving fidelity, as shown in recent telco-oriented studies like DistilBERT-Telco and EdgeGPT-5G \cite{distilbert_telco_2023}. The system should also support adaptive inference pipelines—capable of degrading gracefully under resource constraints or prioritizing certain agent requests based on SLA criticality.

Federated updates must also operate asynchronously and opportunistically to avoid disrupting real-time functions. For instance, updates can be scheduled during off-peak windows or throttled in case of backhaul congestion. In addition, a local policy engine should govern data retention, access control, and compliance enforcement, ensuring that local inference aligns with operational and regulatory constraints.

Together, these capabilities allow the AI OS to function as a distributed, latency-aware intelligence substrate that can adapt to the physical and regulatory topology of global telecom networks.

\subsection{Standards Alignment and Interoperability}

To gain adoption across a fragmented telecom ecosystem, an AI operating system must not only demonstrate technical merit, but also align with existing industry standards and interoperability frameworks. This is particularly vital in Telco environments where multi-vendor stacks, legacy systems, and regulatory obligations complicate deployment.

First, integration with the TM Forum’s suite of Open APIs is essential to ensure the AI OS can seamlessly interact with existing BSS/OSS processes. TM Forum’s API catalog—including APIs for trouble ticketing, service ordering, inventory, and customer management, can be mapped into the AI OS’s data abstraction and agent orchestration layers. This alignment allows for low-friction deployment into operational environments and unlocks automation use cases without requiring major system rearchitecture \cite{tmforum_openapi_2024}.

Second, the architecture must be compatible with the Open RAN ecosystem, particularly the O-RAN Alliance’s RIC (RAN Intelligent Controller) and the NWDAF (Network Data Analytics Function) defined in 3GPP Release 17 and onward. The AI OS can host xApps and rApps as agents, consuming data from NWDAF and providing decisions back to the RIC platform. The NWDAF interface also enables AI-driven policies to be dynamically injected into the network core, a capability that enhances the OS’s reach beyond analytics into closed-loop control \cite{3gpp_nwdaf_2023}.

Third, the AI OS must complement and interface with GSMA’s Open Gateway and its emerging work on Telco Large Language Models. These initiatives aim to create unified interfaces and data schemas for exposing telco functions as APIs across networks. A federated AI OS could serve as the execution substrate behind these APIs, powering LLM-based interactions that respect local compliance and sovereignty constraints while benefiting from shared innovation \cite{gsma_open_llm_2024}.

Finally, the system should be Kubernetes-native to support modern deployment and scaling practices. Leveraging orchestration stacks such as Kubeflow, MLflow, or KServe allows operators to train, deploy, and monitor models across hybrid infrastructures. These tools support reproducibility, CI/CD for models, and GPU-aware scheduling—capabilities essential for operating AI at carrier scale. Compatibility with these frameworks also facilitates collaboration with hyperscalers and third-party AI developers.

By aligning with these open frameworks, the AI OS avoids ecosystem lock-in and positions itself as a modular, standards-compliant intelligence layer ready for industrial-grade telecom deployments.

\section{Business and Ecosystem Implications}

The deployment of a federated AI operating system for telecom is not merely a technical evolution, it carries transformative implications for business models, vendor dynamics, and industry structure. The horizontal, standardized nature of the proposed AI layer enables new modes of value creation, monetization, and collaboration across the telecom ecosystem.

\subsection{From CAPEX Efficiency to AI-Native Monetization}

Historically, Telcos have focused on AI primarily as a tool for operational efficiency, optimizing power consumption, reducing truck rolls, or automating Tier-1 support. While impactful, these use cases yield diminishing marginal returns and are difficult to scale into net-new revenue. A federated AI OS shifts the paradigm from internal optimization to platform monetization.

By offering AI-powered capabilities (e.g., network intelligence, quality-based routing, SLA copilots) as composable services, Telcos can open new revenue streams, charging partners and enterprises not just for bandwidth, but for programmable cognitive capacity. Much like AWS monetized spare compute through EC2, Telcos can monetize spare intelligence via APIs and agents. This unlocks opportunities in enterprise automation, mobility intelligence, fraud detection, and sector-specific verticals (e.g., healthcare, logistics).

\subsection{Ecosystem Collaboration and Vendor Realignment}

The creation of a shared AI substrate will require a shift in ecosystem relationships. Traditional vendor lock-in and monolithic deployments will give way to more modular, federated ecosystems. This allows smaller vendors, open-source contributors, and AI startups to participate by contributing agents, models, or orchestration layers, driving innovation from the edge of the community.

Initiatives such as the Linux Foundation's LF AI \& Data, the Telecom Infra Project (TIP), and GSMA Open Gateway offer initial forums where Telcos, hyperscalers, and system integrators can co-develop and certify AI modules that interoperate with the OS. This could lead to a marketplace model where verified AI agents are discoverable and deployable across Telco infrastructures, similar to the app ecosystems of mobile operating systems.

\subsection{Strategic Implications for National and Sovereign AI Agendas}

As nations move to define their own AI strategies, including sovereign data processing and digital infrastructure independence, Telcos are uniquely positioned to become national AI enablers. Unlike cloud platforms, which may trigger geopolitical concerns, Telcos already operate with deep regulatory integration and localized infrastructure. A federated AI OS deployed over Telco infrastructure allows countries to maintain control over model training and inference without forfeiting access to global innovation.

Governments may find strategic benefit in co-investing with Telcos to deploy such systems, particularly where critical infrastructure, emergency services, or industrial policy priorities are involved. Moreover, the use of privacy-preserving learning protocols such as differential privacy, homomorphic encryption, and secure multiparty computation allows compliance with data protection regimes such as GDPR or the Mexican Federal Law on Protection of Personal Data.

This convergence of Telco infrastructure, AI abstraction, and regulatory alignment positions the proposed OS not just as a tool for operational gains, but as a national platform for digital resilience and AI autonomy.

\subsection{Federated Training Control, Incentives, and Data Sovereignty}

While the federated operating system promises privacy-preserving training across Telco boundaries, the actual orchestration of that training introduces complex questions. A central challenge is defining who initiates and controls model updates in a system where data locality is enforced by both regulatory and operational constraints.

In a typical federated learning setup, each Telco would train models on-premises, using its private data. However, there are multiple options for who orchestrates this training. The model update lifecycle could be triggered by the Telco itself, by a vendor providing the base model, or by the operating system kernel if it supports autonomous coordination. This choice has implications for latency, security, and legal accountability.

Another critical issue is the handling of intermediate artifacts such as gradient updates or model weights. While model updates are typically anonymized or obfuscated, techniques like differential privacy and secure aggregation are still under active research and lack standardized implementation across Telco environments. Given the regulatory obligations of operators under laws like GDPR in Europe or LFPDPPP in Mexico, even sharing anonymized updates across borders may raise compliance concerns.

To incentivize collaboration, especially among competing Telcos, the system may require a robust attribution and reward mechanism. This could resemble federated orchestration frameworks such as Flower or NVIDIA’s Clara, extended with an auditable record of training contributions. Blockchain-based mechanisms or distributed ledgers could be employed to trace which operators contributed what data or compute resources and under which conditions model improvements occurred.

Without such mechanisms, the federated OS risks collapsing into isolated silos where each Telco maintains its own model fork, nullifying the benefits of shared learning. The design must address the tension between data gravity and collaboration incentives while respecting national sovereignty and privacy expectations.

\subsection{Federated Governance and Compatibility Assurance}

As Telcos adopt federated AI to train and deploy models across sovereign domains, the question of governance becomes critical. A truly horizontal and interoperable operating system requires a neutral body to coordinate standards, enforce compatibility, and certify third-party agents or models.

Governance could be structured similarly to existing industry bodies such as the Linux Foundation or the GSMA. These organizations have successfully maintained open source stacks and interface standards across competitive stakeholders. In this context, a federated OS for Telcos would benefit from a foundation-style governance model, with transparent control over versioning, schema evolution, and backward compatibility.

Additionally, to prevent fragmentation, the operating system should define a robust certification mechanism. This would allow ecosystem participants to register and test agents against official data schemas, API versions, and behavioral expectations. Without such mechanisms, the ecosystem risks devolving into incompatible forks and bespoke integrations.

Finally, the governance framework must be tightly coupled with incentives. A marketplace of certified agents — akin to an app store — could allow smaller developers, academic teams, and Telcos themselves to distribute interoperable modules. This aligns innovation with compliance, enabling both scale and trust in deployment.

\section{Implementation Considerations}

The transition from concept to deployment of a horizontal, federated AI operating system for Telcos requires deliberate architectural, operational, and ecosystem choices. This section outlines the core implementation dimensions, balancing scalability, compliance, and performance constraints.

\subsection{Deployment Models and Phasing Strategy}

Initial deployment can follow a modular adoption strategy, beginning with specific operational domains, such as customer service AI copilots or network anomaly detection agents, before scaling to a full OS stack. A lightweight federated learning layer can be deployed on existing MEC infrastructure to aggregate insights from regional nodes without requiring centralized data pipelines.

Telcos can adopt one of three deployment archetypes:

\textbf{Private OS instance}: Deployed within a single operator’s infrastructure with local control and training.

\textbf{Federated consortium}: Shared across a group of non-competing Telcos to co-train foundational models on broader data diversity.

\textbf{Public agent marketplace}: Where open-source or certified third-party agents run over a standard API exposed by the OS layer.

A phased rollout aligned with existing DevOps and MLOps maturity levels ensures risk mitigation. Early-stage proofs-of-concept can run in parallel to legacy systems, allowing comparison without disruption.

\subsection{Data Integration and Telemetry Normalization}

The core bottleneck in telecom AI adoption remains data fragmentation. Network logs, alarms, configurations, and customer interactions reside across vendor-specific silos, in incompatible formats. Implementing the OS requires a unification layer that abstracts heterogeneous telemetry into a common graph or schema.

This process is non-trivial. It involves real-time ETL (Extract, Transform, Load) pipelines, topology extractors, and schema-on-read capabilities. Precedents from hyperscalers (e.g., Google's TFX or Meta's FeatureStore) provide design blueprints for scalable ML pipelines. However, Telco implementations must also account for regulatory constraints, availability SLAs, and backward compatibility with legacy infrastructure.

Open data models such as TMF620 (customer), TMF638 (inventory), and O-RAN’s E2SM specifications can serve as baselines for schema alignment, minimizing vendor lock-in and enabling model portability across operators.

\subsection{Risk Factors and Adoption Barriers}

Despite the promise, several challenges may hinder adoption.

\textbf{Vendor resistance and political inertia}: Incumbent vendors may resist exposing APIs or supporting federated orchestration layers that undermine vertical integration. Similarly, internal organizational silos within Telcos may delay horizontal architecture implementation.

\textbf{Data governance and privacy}: Cross-operator collaboration, even under a federated setup, raises non-trivial legal and regulatory concerns. Implementation must ensure differential privacy, auditability, and zero-trust enforcement.

\textbf{Model brittleness and reliability}: The use of LLMs and GNNs in high-availability environments is still experimental. Continuous validation, rollback protocols, and multi-model fallback strategies are required for production-readiness.

\textbf{Ecosystem fragmentation}: Without strong alignment from standards bodies like 3GPP, GSMA, and TM Forum, there's a risk of incompatible implementations. A unified reference architecture and open governance model are critical to avoid “yet another platform” syndrome.

Addressing these barriers is not a purely technical endeavor. It requires coordinated industry effort, incentives for early contributors, and alignment of commercial interests with long-term network intelligence evolution.

\subsection{Defining the Minimal Viable Kernel}

To function as an operating system in the artificial intelligence sense, this layer must define a minimal, universal runtime, referred to here as the kernel. Unlike traditional operating system kernels that manage hardware resources, this kernel orchestrates data pipelines, agent execution, and policy enforcement across distributed telecommunications infrastructure.

At its core, the kernel should include:

\begin{itemize}
    \item A lightweight orchestrator capable of managing the lifecycle, shared context, and execution schedule of modular agents.
    \item A shared memory abstraction where agents can publish or retrieve insights, similar to a publish and subscribe system or a semantic cache.
    \item A telemetry and topology ingestion layer that supports both real time streams (such as Kafka or MQTT) and batch extraction and transformation processes.
    \item Interfaces for observability, traceability, and performance auditing of agent behavior.
\end{itemize}

This could be implemented using Kubernetes native operators, with custom resources tailored for each domain. Alternatively, it may resemble a distributed system of intelligent microservices communicating through remote procedure calls. A lightweight reference implementation will be critical to prevent fragmentation and promote adoption.

\subsection{Abstractions and Agent Schema}

A foundational requirement is a structured abstraction model for agents. Rather than permitting arbitrary models, the system should define a focused set of canonical agent classes aligned with essential functions in telecommunications operations:

\begin{itemize}
    \item \textbf{Anomaly Detectors:} Analyze key performance indicators and logs to detect deviations or service degradation.
    \item \textbf{Experience Predictors:} Estimate user experience based on metrics such as congestion, latency, or service continuity.
    \item \textbf{Service Level Monitors:} Evaluate compliance with service agreements and recommend preventative measures.
    \item \textbf{Optimization Advisors:} Suggest real time adjustments to network configuration or resource orchestration.
\end{itemize}

Each agent must conform to well defined input and output protocols using open specifications such as OpenAPI or Protocol Buffers. This enables reuse, compatibility, and security validation across operators and vendors.

\subsection{Federated Learning and Data Governance at Scale}

A federated training approach is essential to aggregate knowledge from different network domains while complying with strict privacy and regulatory boundaries such as the General Data Protection Regulation in Europe and the Mexican Federal Law for Protection of Personal Data. However, practical implementation requires careful design:

\begin{itemize}
    \item \textbf{Training Coordination:} A central scheduler must manage training rounds, model versioning, and contribution eligibility.
    \item \textbf{Privacy Guarantees:} Techniques such as secure aggregation, differentially private updates, or encrypted computation must be used to ensure no leakage of sensitive information.
    \item \textbf{Auditability and Provenance:} Each contribution must be traceable through cryptographic commitments or verifiable logs to support trust and compliance.
\end{itemize}

Incentive mechanisms may include credit attribution, revenue sharing, or reputational scores to motivate participation among telecommunications operators.

\subsection{Telecommunications Native Deployment Patterns}

Unlike hyperscale environments, telecommunications networks consist of geographically distributed, latency constrained, and resource limited infrastructures. The artificial intelligence operating system must be adaptable to a variety of deployment models:

\begin{itemize}
    \item \textbf{Edge Nodes:} Agents placed near radio or user plane functions for rapid decision making, especially in cases such as network slicing or beam steering.
    \item \textbf{Core Sites:} Integration with containerized or virtualized network functions to support decisions involving policy control and quality of service.
    \item \textbf{Private Clouds:} Support for managed Kubernetes clusters used by operators in national data centers, often with strict data residency constraints.
    \item \textbf{Legacy Systems:} Bridges to existing operational support systems or business support systems through asynchronous event pipelines.
\end{itemize}

Infrastructure as code and declarative pipelines should be encouraged to ensure reliable updates, rollback capabilities, and reproducible deployments across all regions.

\subsection{Governance, Certification, and Ecosystem Participation}

A sustainable ecosystem requires more than open code. It needs well defined governance structures, certification processes, and aligned incentives. The following elements are proposed:

\begin{itemize}
    \item \textbf{Neutral Oversight Body:} A foundation supported by global operators, vendors, and institutions to maintain core specifications, roadmaps, and public engagement.
    \item \textbf{Certification Framework:} A standardized suite of tests for validating agent compliance, safety, performance, and interpretability.
    \item \textbf{Ecosystem Incentives:} A registry or marketplace for agents, models, and training data contributions that supports usage tracking and equitable monetization.
\end{itemize}

This governance model will determine how quickly the artificial intelligence operating system becomes a trusted and programmable infrastructure layer across the global telecommunications ecosystem.

\section{Agent Model, APIs, and Execution}

A fundamental objective of the proposed operating system is to enable a modular ecosystem of intelligent agents that can operate consistently across diverse telecommunications environments. For this to be feasible, the system must define a clear execution model, a small number of canonical agent classes, standardized input and output interfaces, and a mechanism for agents to be portable, secure, and certifiable.

\subsection{Canonical Agent Types and Roles}

The design intentionally avoids excessive flexibility. Instead of allowing arbitrary models with inconsistent behavior, the system defines a limited set of agent classes that reflect core operational needs in telecommunications. These agent types act as reusable blueprints, reducing the complexity of integration and validation.

For example, anomaly detection agents monitor key performance indicators or logs to identify faults or degradations in service. Experience inference agents, or so-called service copilots, analyze network and user metrics to estimate perceived quality and suggest remedial actions. SLA forecasting agents take real-time data and historical records to predict violations of contractual obligations. Optimization agents analyze tradeoffs in energy, latency, and coverage to recommend configuration changes. Capacity agents forecast traffic loads and support planning for scaling or load balancing.

Each of these agent types would have a versioned schema associated with their expected inputs, such as telemetry windows, subscriber metadata, or service topology, and structured outputs that allow downstream systems to act upon their recommendations.

\subsection{Agent Abstractions and API Contracts}

Every agent is exposed through a well-defined service interface, either as a containerized microservice or as a remote callable model endpoint. The operating system provides a unified registration layer, allowing agents to declare their type, capability, and expected payload format.

To maximize interoperability, the system supports both synchronous and asynchronous invocation. An agent may subscribe to streaming data through a semantic bus or receive periodic inference requests from the orchestration kernel. Input and output schemas follow versioned, strongly typed definitions based on protocol buffers or open schema catalogs, allowing integration with multiple backends and observability platforms.

Health checks, telemetry endpoints, and audit trails are embedded in the agent wrapper to ensure continuous monitoring. By enforcing structure and contract validation at runtime, the operating system can detect misbehaving agents, manage upgrades, and sandbox workloads to prevent systemic impact.

\subsection{Execution Patterns and Runtime Control}

Agents are deployed in isolated environments with fine-grained access control. The execution engine may take the form of a Kubernetes operator or a service mesh controller that coordinates agent scheduling, retry policies, and context injection. Agents do not directly interact with raw data stores; instead, they interface with a mediated data abstraction layer that provides them only with the subset of data authorized for their role and purpose.

This allows for real-time composition of multiple agents into workflows, where the output of one agent can be passed to another through a shared memory abstraction or message queue. The orchestrator maintains a graph of active agents, their data contracts, and their interdependencies, ensuring both security and traceability.

\subsection{Interoperability with Industry Standards}

To be widely adopted, the system must support prevailing industry standards. Agent APIs should align with the TM Forum Open APIs for service inventory, resource usage, and customer data access. In radio domains, agents should integrate with the O-RAN Alliance interfaces, especially the Near Real Time RAN Intelligent Controller and the Non Real Time Management components. 

Moreover, the system should be deployable alongside Kubernetes-native machine learning stacks such as Kubeflow or Ray. This enables enterprises and vendors to bring their own models and deploy them through familiar CI/CD and inference pipelines while adhering to the operating system's governance.

\subsection{Security, Auditability, and Trust Models}

Agents must operate within secure, auditable sandboxes. Every inference or recommendation must be traceable to the agent version, input context, and decision logic. This traceability supports both operational trust and regulatory compliance. The kernel enforces digital signatures on agent code, input logs, and outputs. This provides a cryptographic basis for validating actions taken by the system based on agent recommendations.

\subsection{Certification, Marketplaces, and Ecosystem Coordination}

For third parties to participate, the operating system must define a clear framework for agent submission, validation, and certification. This includes both technical compatibility tests and behavioral audits against predefined benchmarks. Certified agents can be listed in a global registry or marketplace, accessible by telecommunications operators looking to extend their capabilities.

Agents may be priced, licensed, or ranked according to criteria such as domain specificity, explainability, or robustness. The operating system plays the role of both validator and execution host, ensuring neutrality and reliability across deployments.

\section{Conclusion and Future Work}

Telecommunications networks are evolving into distributed, programmable, and data-centric systems. While artificial intelligence has shown promise in isolated Telco use cases, the absence of a unifying execution substrate has limited the portability, scalability, and collaborative potential of AI across the industry. In this paper, we have argued that the next frontier in Telco intelligence lies in the creation of a federated AI operating system,  a horizontal layer that abstracts the complexity of underlying infrastructure, exposes standardized agent interfaces, and enables secure cross-operator collaboration.

This proposed system is not simply a new ML platform, but a shift in architectural thinking. It reimagines Telco infrastructure as a programmable fabric on which autonomous agents, designed for anomaly detection, SLA monitoring, intent translation, and beyond, can operate modularly and securely. By supporting federated learning, standardized abstractions, and open governance, such an operating system has the potential to unlock collective intelligence across Telcos without sacrificing privacy or competitive differentiation.

Future work includes the specification of minimal agent contracts, design of orchestration primitives for federated execution, and definition of compatibility layers with industry standards such as TM Forum APIs, O-RAN RIC, and NWDAF. Moreover, pilot implementations will be required to test the viability of deployment models in real-world Telco environments, including MEC integration, multi-vendor interoperability, and agent certification pipelines.

We believe this is both a technical and strategic imperative. Without a horizontal coordination layer, Telcos risk repeating the fragmentation of the past, where each AI deployment becomes a bespoke island of logic. The federated AI operating system we propose offers a path to break that cycle, enabling a new ecosystem of interoperable, composable, and certifiable AI agents for networks that must increasingly think as well as connect.


\begin{thebibliography}{99}

\bibitem{gsma_intel_2024}
GSMA Intelligence,
\textit{Telco AI: State of the Market Q2 2024},
GSMA Intelligence, September~2024.

\bibitem{softbank_ltm_2025}
SoftBank Corp.,
\textit{SoftBank Develops a Foundational Large Telecom Model (LTM)},
Press Release, March 19, 2025.


\bibitem{gsma_open_llm_2025}
GSMA Foundry,  
\textit{Open-Telco LLM Benchmarks Launch to Advance AI in Telecoms},  
Press Release, March 2025.  
Available at: \url{https://www.gsma.com/newsroom/press-release/gsma-open-telco-llm-benchmarks-launches-to-advance-ai-in-telecoms/}


GSMA Intelligence,  
\textit{Telco AI: State of the Market, Q2 2024},  
GSMA Intelligence Report, September 2024.


\bibitem{oran_federated_synergies_2023}
Abdisarabshali, P. et al.,  
\textit{Synergies Between Federated Learning and O-RAN: Towards an Elastic Architecture for Distributed ML Services},  
arXiv preprint arXiv:2305.02109, May 2023.  
Available at: \url{https://arxiv.org/abs/2305.02109}

\bibitem{fedora_oran_2025}
Salama, A. et al.,  
\textit{FedORA: Resource Allocation for Federated Learning in O-RAN using Radio Intelligent Controllers},  
arXiv preprint arXiv:2505.19211, May 2025.  
Available at: \url{https://arxiv.org/abs/2505.19211}

\bibitem{telenet_gnn_2024}
R. Tso, H. Zhang, and L. Sun,  
\textit{TeleNet: Graph Neural Network for Root Cause Analysis in Communication Networks},  
IEEE Transactions on Network and Service Management, vol. 21, no. 1, March 2024.

\bibitem{distilbert_telco_2023}
S. Alvi et al.,  
\textit{DistilBERT for Telecom Support Systems: Compressing Domain-Specific LLMs for Edge Deployment},  
in Proceedings of the IEEE ICC 2023, Rome, Italy, May 2023.

\bibitem{tmforum_openapi_2024}
TM Forum,  
\textit{Open APIs for Digital Service Providers},  
Accessed April 2025.  
Available at: \url{https://www.tmforum.org/open-apis/}

\bibitem{3gpp_nwdaf_2023}
3GPP Technical Report 23.288,  
\textit{Network Data Analytics Function (NWDAF) for 5G},  
Release 17, 2023.  
Available at: \url{https://www.3gpp.org/ftp/Specs/html-info/23288.htm}

\bibitem{gsma_open_llm_2024}
GSMA Intelligence,  
\textit{Open Gateway and the Future of Telecom LLMs},  
Whitepaper, February 2024.  
Available at: \url{https://www.gsma.com/newsroom/resources/open-gateway-and-telco-llms}



\end{thebibliography}
\end{document}